\documentclass[floatfix,aps,showpacs,twocolumn,prb,
,superscriptaddress]{revtex4}%
\usepackage{amsmath}
\usepackage{amssymb}
\usepackage{graphicx}
\usepackage{dcolumn}
\usepackage{natbib}
\usepackage{bm}
\usepackage{epsfig}
\usepackage{amsfonts}%
\providecommand{\U}[1]{\protect\rule{.1in}{.1in}}
\providecommand{\U}[1]{\protect\rule{.1in}{.1in}}
\providecommand{\U}[1]{\protect\rule{.1in}{.1in}}
\providecommand{\U}[1]{\protect\rule{.1in}{.1in}}

\begin{document}
\title{Spin polarized current in a junction of zigzag carbon nanotube}
\author{Zhan-Feng Jiang}
\affiliation{Beijing National Laboratory for Condensed Matter Physics, Institute of
Physics, Chinese Academy of Sciences, Beijing 100080, China }
\affiliation{Department of Physics, and Center of Theoretical and Computational Physics,
The University of Hong Kong, Hong Kong}
\author{Jian Li}
\affiliation{Department of Physics, and Center of Theoretical and Computational Physics,
The University of Hong Kong, Hong Kong}
\author{Shun-Qing Shen}
\affiliation{Department of Physics, and Center of Theoretical and Computational Physics,
The University of Hong Kong, Hong Kong}
\author{W. M. Liu}
\affiliation{Beijing National Laboratory for Condensed Matter Physics, Institute of
Physics, Chinese Academy of Sciences, Beijing 100080, China }
\date{\today}

\begin{abstract}
We investigated spin-resolved electronic transport through a junction composed
of a nonmagnetic metal electrode and a zigzag carbon nanotube (ZCNT) by means
of self-consistent Green's function method in the tight binding approximation
and the unrestricted Hartree-Fock approximation. Our results show that the
electric current can be spin-polarized if the coupling of the junction is
weak. Further calculations on spin-spin correlation and local density of
states reveal the existence of magnetic edge states in ZCNTs, which is
responsible for the observed spin-polarized current and can be controlled by
applying a gate voltage. We also studied the influence of the nearest-neighbor
Coulomb interaction and the junction coupling strength on the
spin-polarization of the current.

\end{abstract}

\maketitle

\section{INTRODUCTION}

Carbon nanotubes (CNTs) have been the subject of a great amount of theoretical
and experimental works during the past two decades, because of their rich and
fascinating properties and potential applications in nanotechnology.
Electronic structures and transport properties of infinite-length CNTs have
been extensively studied and well understood.\cite{infinite,infinitebook} A
finite-length CNT, on the other hand, is usually treated only as a simplified
one-dimensional quantum wire in transport theories,\cite{arXiv} or tunneling
experiments,\cite{ex1,ex2,ex3,ex4,ex5,ex6} with its crystal fine structures
largely ignored. In fact, the hexagonal lattice structure of a CNT, inherited
from the graphene lattice, may give rise to peculiar properties near its
edges. One example is a recent work by Son \textit{et al.}, where a zigzag
graphene nanoribbon is predicted to be half-metallic when an strong-enough
external electric field is applied transversely, due to its spin-polarized
edge states.\cite{stevennature} Similarly the ground state of a ZCNT also
shows antiferromagnetic order as a result of the Coulomb interaction and
correlation of electrons in the hexagonal crystal lattice. This is explained
by two mathematical theorems for the Hubbard model defined on a bipartite
lattice with equal numbers of sublattice sites, \textit{i.e.}, the ground
state is a spin singlet\cite{lieb1} and the antiferromagnetic correlation is
always dominant over other correlations.\cite{lieb2} Some literatures have
investigated the band structures and antiferromagnetic orders (or spin density
waves) in graphene ribbons or CNTs by the extended Hubbard
Model.\cite{jap1,jap2,jap3,UV1,UV2,HF1,HF2,wave1,wave2} Few works have been
done, however, on the edge states of CNTs and their effects on transport.

In fact, not only zigzag graphene nanoribbons but also finite-length ZCNTs
have spin-polarized edge states due to the Coulomb interaction and their
bipartite lattice structures.\cite{Lin06} The spin polarized edge states
existing in finite or semi-infinite ZCNTs can play an important role in their
spin transport properties. Particularly, in a junction composed of a
semi-infinite nonmagnetic metal (NM) electrode and a semi-infinite ZCNT, an
edge state may exist in the ZCNT near the interface. The edge state is spin
polarized, with oscillations upon different sublattices, and forms a
spin-dependent scattering potential. It will act as a spin filter when
electrons tunnel through the interface to induce a spin-polarized current. In
this paper, we use the extended Hubbard model to investigate the
spin-dependent transport of the junction of the ZCNT to the NM lead.

The paper is arranged as follows. In Sec.II we introduce the model Hamiltonian
and describe the general formalism of our self-consistent method. In Sec.III,
the main results and discussions are presented. Sec.IV is a brief summary, and
finally, an appendix is attached to discuss the validity of the
self-consistent method we used in this paper.

\section{General Formalism}

\subsection{Model Hamiltonian}

\begin{figure}
[ptb]
\begin{center}
\includegraphics[
height=2.2494in,
width=2.8911in
]%
{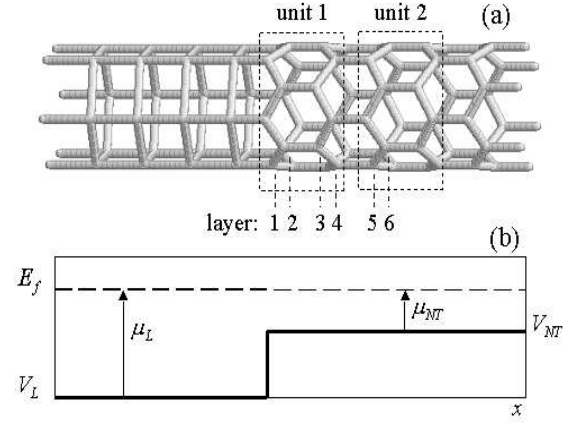}%
\caption{(a) Schematic of a NM-ZCNT junction, the dashed rectangles denote the
unit cells of ZCNT, and the dashed lines denote the layers of ZCNT. (b)
Schematic diagram of the electrostatic potential and electrochemical
potential. }%
\label{graph1}%
\end{center}
\end{figure}

We consider a conjunction of a semi-infinite tubal lead and a semi-infinite
ZCNT, each of which extends to an electron reservoir at infinity, as shown in
Fig. \ref{graph1}(a). The ZCNT part is the standard hexagonal-crystal tube,
and the NM part is assumed to be a square-lattice tube, which is connected to
the terminal layer (layer-1 in Fig. \ref{graph1}(a)) of the ZCNT. In the
tight-binding approximation, the model Hamiltonian for the whole system reads%

\begin{equation}
H=H_{NM}^{L}+H_{int}^{L}+H_{NT}, \label{hall}%
\end{equation}
where
\begin{subequations}
\label{hi}%
\begin{align}
H_{NM}^{L}  &  =-\underset{\left\langle i,j\right\rangle ,\sigma}{\sum}%
t_{L}d_{i\sigma}^{+}d_{j\sigma}+\underset{i,\sigma}{\sum}V_{L}d_{i\sigma}%
^{+}d_{i\sigma},\\
H_{int}^{L}  &  =-\underset{\left\langle i,j\right\rangle ,\sigma}{\sum
}t_{int}^{L}d_{i\sigma}^{+}c_{j\sigma}+h.c.,\\
H_{NT}  &  =-\underset{\left\langle i,j\right\rangle ,\sigma}{\sum}%
tc_{i\sigma}^{+}c_{j\sigma}+U\underset{i}{%
{\displaystyle\sum}
}n_{i\uparrow}n_{i\downarrow}\nonumber\\
&  +V\underset{\left\langle i,j\right\rangle }{%
{\displaystyle\sum}
}n_{i}n_{j}+V_{NT}\underset{i,\sigma}{%
{\displaystyle\sum}
}n_{i\sigma}. \label{HNT}%
\end{align}
Here, $H_{NM}^{L}$ describes the semi-infinite NM lead, and $d_{i\sigma}%
^{+}(d_{i\sigma})$ is the creation (annihilation) operator of an electron at
the $i^{th}$ site in the NM part with spin-$\sigma$ $(\sigma=\uparrow
,\downarrow)$. $\left\langle i,j\right\rangle $ denotes a pair of
nearest-neighboring sites $i$ and $j$ in the lattice, and $t_{L}$ represents
the hopping integral between them, and $V_{L}$ is the electrostatic potential
in the NM part. $H_{int}^{L}$ represents the tunneling process between the NM
lead and the ZCNT. In Fig. \ref{graph1}(a), $t_{int}^{L}\neq0$ connects the
interface layers of the NM lead and the ZCNT. $H_{NT}$ describes the $\pi
$-orbital electrons of the zigzag nanotube. We adopt the extended Hubbard
model including the on-site Coulomb interaction $U$ and nearest-neighbor
interaction $V$.\cite{HF1,HF2,HF3} $c_{i\sigma}^{+}(c_{i\sigma})$ is the
creation (annihilation) operator of an electron at the $i^{th}$ site in the
ZCNT with spin-$\sigma$, $n_{i\sigma}=c_{i\sigma}^{+}c_{i\sigma}$, and $t$ is
the hopping integral, which is about 2.7eV for carbon
nanotubes,\cite{stevenprl} taking the unit of energy in this paper. $V_{NT}$
is the electrostatic potential in the ZCNT.

To solve the interacting problem, we perform the unrestricted Hartree-Fock
approximation, and reduce $H_{NT}$ to a mean-field one,%

\end{subequations}
\begin{align}
h_{NT} &  =-\underset{\left\langle i,j\right\rangle ,\sigma}{\sum
}(t+V\left\langle c_{j\sigma}^{+}c_{i\sigma}\right\rangle )c_{i\sigma}%
^{+}c_{j\sigma}+V_{NT}\underset{i,\sigma}{%
{\displaystyle\sum}
}n_{i\sigma}\nonumber\\
&  +U\underset{i}{%
{\displaystyle\sum}
}(\left\langle n_{i\uparrow}\right\rangle n_{i\downarrow}+\left\langle
n_{i\downarrow}\right\rangle n_{i\uparrow})+V\underset{\left\langle
i,j\right\rangle }{%
{\displaystyle\sum}
}\left\langle n_{i}\right\rangle n_{j},\label{hh}%
\end{align}
where the expectation values of physical quantities $\left\langle \cdot
\cdot\cdot\right\rangle $ need to be calculated self-consistently. We use the
zero-temperature Green's function method to establish a set of mean field
equations, and calculate these expectation values for the system
self-consistently. The technique of principal layers is used to solve the
Green's function of a semi-infinite lead.\cite{self1,self2} This technique,
limited to treat periodic structures, is suitable for the NM lead, but not for
the ZCNT part, because the inhomogeneity of $\left\langle c_{j\sigma}%
^{+}c_{i\sigma}\right\rangle $ and $\left\langle n_{i\sigma}\right\rangle $
near the interface of the ZCNT will make the mean-field Hamiltonian $h_{NT}$
generally not a periodic one and thus we cannot apply this technique directly.
But we notice that the expectation values deep inside the ZCNT part are nearly
periodic and very close to the corresponding values in an infinite ZCNT. So we
divide the ZCNT into two part: the central part with N units (see Fig.
\ref{graph1}(a) for the definition of 'unit' in this paper) including the
interface, and the semi-infinite right part. We assume the expectation values
in the right part are periodic, in this way the Green's function of the right
part can now be solved with the technique of principal layers and the whole
system can be solved self-consistently. Thus although the whole system in Fig.
\ref{graph1} has only two compositions, in practice we treat it in three
parts: the NM part on the left naturally, N units of the ZCNT including the
interface at the center, and the rest of the ZCNT on the right. The assumption
that the expectation values $\left\langle c_{j\sigma}^{+}c_{i\sigma
}\right\rangle $ and $\left\langle n_{i\sigma}\right\rangle $ are periodic in
the right part, and equal to those in an otherwise integral ZCNT, is justified
when $N$ is so large that the expectation value of any quantity in the ZCNT
changes little with increased $N$. In Appendix we will discuss this issue in
more details. Based on the above analysis, the Hamiltonian for the ZCNT can be
rewritten as
\begin{equation}
h_{NT}=h_{NT}^{M}+h_{int}^{R}+h_{NT}^{R},\label{htide}%
\end{equation}
where $h_{NT}^{M}$ and $h_{NT}^{R}$ describe the central part and the
semi-infinite right part, respectively, $h_{int}^{R}=-\underset{\left\langle
i,j\right\rangle ,\sigma}{\sum}(t+V\left\langle c_{j\sigma}^{+}c_{i\sigma
}\right\rangle )c_{i\sigma}^{+}c_{j\sigma}$ represents the connection between
the central part $h_{NT}^{M}$ and the right part $h_{NT}^{R}$. In this way,
$h_{NT}^{R}$ can be treated by the technique of principal layers.

In order to reduce the influence of band structure of the nonmagnetic metal
lead on the transport property, we take $t_{L}=4t$, which is equivalent to the
so-called \textquotedblleft wide-band limit\textquotedblright.\cite{wideband}
And in order to preserve the edge state of the ZCNT, we take $t_{int}%
^{L}=0.1t$, which is a comparatively small value implying weak coupling
between the NM lead and the ZCNT, or a high potential barrier between them.
This can be realized in experiments by fabricating an oxide layer at the
interface.\cite{arXiv} In equilibrium, the Fermi level of the whole system
$E_{f}$ is identical everywhere as shown in Fig. \ref{graph1}(b). Following
the definitions in Ref. [\onlinecite{potential}], the electrochemical
potential (ECP) in the ZCNT is defined by $\mu_{NT}\equiv E_{f}-V_{NT}$, and
the electrochemical potential in the NM part is defined by $\mu_{L}\equiv
E_{f}-V_{L}$. In this paper, $E_{f}$, $V_{L}$, and $\mu_{L}$ are fixed,
$V_{NT}$ is assumed to be adjustable by means of a gate voltage. Adjusting
$\mu_{NT}$ may control the electron density $n_{e}$ in the ZCNT. We will fix
the NM part at half-filling by tuning $V_{L}$ (or $\mu_{L}$), so that the
transport modes in the left lead around the Fermi level can match those in the
ZCNT, to facilitate the electronic transport.\cite{nano}

\subsection{Method}

We come to establish the mean field equation by means of the self-consistent
Green's function method, and then to use the recursive method to find the
numerical solution. The retarded Green's function is defined by
\begin{equation}
G_{i\sigma,j\sigma^{\prime}}^{r}(t;t^{\prime})=-i\theta(t-t^{\prime
})\left\langle \left\{  c_{i\sigma}(t),c_{j\sigma}^{+}(t^{\prime})\right\}
\right\rangle ,
\end{equation}
whose Fourier transform can be calculated by the Dyson's Equation\cite{datta}%
\begin{equation}
G^{r}(E)=\frac{1}{E+i\eta-h_{NT}^{M}-\Sigma_{L}(E)-\Sigma_{R}(E)},
\label{retard}%
\end{equation}
where $\eta$ is an infinitesimal constant, and $h_{NT}^{M}$ is the mean-field
Hamiltonian of the central part as defined in the last subsection. $\Sigma
_{L}=H_{int}^{L+}g_{L}^{r}H_{int}^{L}$ is the self-energy due to the left
lead, where $H_{int}^{L+}$ is the Hermitian conjugate of $H_{int}^{L}$ and
$g_{L}^{r}=(E+i\eta-H_{NM}^{L})^{-1}$ is the retarded Green's function of the
left lead. The self-energy due to the right lead has the similar definition
$\Sigma_{R}=h_{int}^{R+}g_{R}^{r}h_{int}^{R}$ and $g_{R}^{r}=(E+i\eta
-h_{NT}^{R})^{-1}$. In the present approach, the system is defined on a
lattice and the Hamiltonian $h_{NT}^{M}$ as well as the self energy can be
re-expressed in the form of square matrix. The dimensionality of the matrix is
twice of the lattice number of the cental part of carbon nanotube because the
spin degree of freedom. For a finite number of lattice sites, the Green's
function can be calculated numerically once the initial values are assigned to
the expected quantities in Eq.(\ref{hh}) such as the local densities of charge
and spin. Once we have this Green's function, the expectation value of any
physical quantity can be obtain as follows
\begin{equation}
\left\langle c_{i\sigma}^{+}c_{j\sigma^{\prime}}\right\rangle =\frac{1}{2\pi
i}%
{\displaystyle\int_{-\infty}^{E_{f}}}
dE\cdot\lbrack G_{i\sigma,j\sigma^{\prime}}^{r\ast}(E)-G_{j\sigma^{\prime
},i\sigma}^{r}(E)], \label{lesser}%
\end{equation}
where $G_{i\sigma,j\sigma^{\prime}}^{r\ast}(E)$ is the complex conjugate of
$G_{i\sigma,j\sigma^{\prime}}^{r}(E)$. Then the local charge and spin density
are given by
\begin{equation}
n_{e}(i)=\left\langle c_{i\uparrow}^{+}c_{i\uparrow}+c_{i\downarrow}%
^{+}c_{i\downarrow}\right\rangle , \label{ne}%
\end{equation}
and
\begin{equation}
S(i)=\left\langle c_{i\uparrow}^{+}c_{i\uparrow}-c_{i\downarrow}%
^{+}c_{i\downarrow}\right\rangle ,
\end{equation}
respectively.

The recursive method is applied to obtain the stable solution for the problem.
The stable solutions of the Green's functions are applied to calculate the
relevant physical quantities. The spin polarization (SP) of the conductance is
defined by
\begin{equation}
S_{P}\equiv\frac{G_{\uparrow}-G_{\downarrow}}{G_{\uparrow}+G_{\downarrow}},
\label{spinpola}%
\end{equation}
where $G_{\sigma}$ is the conductance in the spin-$\sigma$ channel
($\sigma=\uparrow,\downarrow$), which is given by the Landauer's formula,
\begin{equation}
G_{\sigma}=\frac{e^{2}}{h}Tr[\Gamma_{L\sigma}G^{r}\Gamma_{R\sigma}%
G^{r+}]_{E=E_{f}}, \label{Gsigma}%
\end{equation}
where $\Gamma_{A\sigma}\equiv i(\Sigma_{A\sigma}-\Sigma_{A\sigma}^{+})$, and
$\Sigma_{A\sigma}$ represents the self-energy of lead $A$ $(=L,R)$ in the
spin-$\sigma$ subspace, $G^{r+}$ is the Hermitian conjugate of $G^{r}$.

\begin{figure}
[ptb]
\begin{center}
\includegraphics[
height=3.0467in,
width=3.1185in
]%
{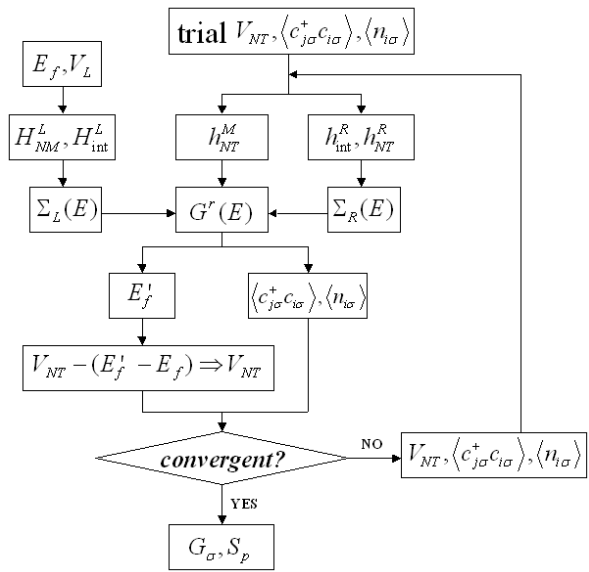}%
\caption{The flow chart of the self-consistent calculation.}%
\label{tree}%
\end{center}
\end{figure}

For concreteness, we describe the overall procedure of the present
self-consistent calculation, which is illustrated in a flow chart in Fig.
\ref{tree} and elucidated as follows:

Step 1: Calculate the self-energy due to the left lead. Because there's no
Coulomb interaction term in the Hamiltonian of the left lead, we can solve
this part without recursion. An arbitrary value can be assigned to $V_{NM}$ to
complete the Hamiltonian $H_{NM}^{L}$,\ then $g_{L}^{r}(E)$ and $\Sigma
_{L}(E)$ can be calculated by the technique of principal
layer\cite{self1,self2}. It's found that the left lead is half-filled when the
Fermi energy $E_{f}=V_{L}$, as the left lead is a simple square lattice. The
Fermi level $E_{f}$ is identical for the whole system because we are studying
in the linear response regime, and it is kept unchanged in the following steps.

Step 2: Construct a trial Hamiltonian $h_{NT}$. An initial charge and spin
distribution is assigned in the zigzag nanotube. The initial change density
promises that the number of electrons per site is about 1, which represents
the half-filled case of the zigzag nanotube. Spin density wave and charge
density wave are constructed on the zigzag nanotube at the same time, because
many literatures have referred to the coexistence of the spin density wave and
charge density wave in carbon nanotubes or ribbons\cite{HF1,HF2,wave1,wave2},
and this is also a general property of the half-filled extended Hubbard
model\cite{wave3,wave4}. The hopping term $\left\langle c_{j\sigma}%
^{+}c_{i\sigma}\right\rangle $ is assigned to be 0, because we have no
knowledge about it so far. Remember $\left\langle c_{j\sigma}^{+}c_{i\sigma
}\right\rangle $ and $\left\langle n_{i\sigma}\right\rangle $ in the right
unit of the central part is also used to determine the Hamiltonian
$h_{int}^{R}$ and $h_{NT}^{R}$. The electrostatic potential $V_{NT}$ can be
assigned the same value with $E_{f}$ at first, because all cases are nearly
half-filled. $V_{NT}$ will be changed in the following iterations. Although
$V_{NT}$ is fixed by the gate voltage in experiments, we cannot use it to be
the restrictive condition in the calculation, otherwise the iterations won't
converge. We use the local charge density $n_{e}$ in the right lead to be the
restrictive condition, which can lead to convergence. Because $n_{e}$ has a
one-to-one correspondance to $V_{NT}$, $V_{NT}$ will be determined along with
the procedure of convergence.

Step 3: Calculate the Green's function $G^{r}(E)$ in Eq.(\ref{retard}). The
self-energy $\Sigma_{R}(E)$ can be obtained from the Hamiltonian $h_{int}^{R}$
and $h_{NT}^{R}$.

Step 4: Calculate the expectation values of the physical quantities
$\left\langle c_{j\sigma}^{+}c_{i\sigma}\right\rangle $ and $\left\langle
n_{i\sigma}\right\rangle $ from Eq.(\ref{lesser}) and (\ref{ne}), and find a
new Fermi energy $E_{f}^{\prime}$. The restrictive condition for finding
$E_{f}^{\prime}$ is the local charge density $n_{e}$ in the right lead is a
certain value. After that we have to draw the Fermi energy back to the initial
value $E_{f}$ by means of tuning $V_{NT}$. So we substitute $V_{NT}%
-(E_{f}^{\prime}-E_{f})$ for $V_{NT}$.

Step 5: Judge the convergence. If the expectation values of the physical
quantities $\left\langle c_{j\sigma}^{+}c_{i\sigma}\right\rangle $ and
$\left\langle n_{i\sigma}\right\rangle $ has converged, we can end the
iterations and get the stable solutions. Otherwise we have to use $V_{NT}$,
$\left\langle c_{j\sigma}^{+}c_{i\sigma}\right\rangle $ and $\left\langle
n_{i\sigma}\right\rangle $ to update the Hamiltonian $h_{NT}$, then turn to
step 3 to begin a new iteration.

Step 6: After convergence, we use the the stable solutions of the Green's
functions to calculate the spin polarization of conductance by
Eq.(\ref{spinpola}) and (\ref{Gsigma}).%

When we calculate the expectation values from Eq.(\ref{lesser}), for the sake
of precision, we employ the contour integration in the complex energy
plane\cite{guohong}, instead of direct integration along the real energy axis,
because the retarded Green's function can be analytically continued into the
complex energy plane. On the integration path, the complex energy has a large
imaginary part, so the infinitesimal constant $\eta$ in Eq.(\ref{retard})
becomes negligible, even can be assigned 0. But when we calculate
Eq.(\ref{Gsigma}), the constant $\eta$ becomes indispensable. It should be
assigned a value as small as possible, so that the conductances are not
influenced by it.

On the magnitudes of $U$ and $V$, $U$ was adopted in a range of $2t\sim4t$ and
$V\approx t$ for graphene, ribbons or nanotubes in some
literatures\cite{UV1,HF2,HF3}. They were estimated from some organic molecules
or by fitting the exciton energies of nanotubes. People also try to use the
first principle calculations to obtain the values of $U$ and $V$. But as the
authors of the Ref. [\onlinecite{wave1}] pointed out, the values of these
parameters depend on the choice of the exchange-correlation functional used
within the calculation of density-functional theory. For example, $U$ is
chosen at $0.9t\sim2t$ when one studied the problem of edge states\cite{wave1}%
. Here we adopt this result to discuss the effect of $U$ and $V$ in several cases.

\section{Results and Discussion}

\subsection{The case of $V=0$}

In this section, we first consider the case of $V=0$ and focus on the effect
of $U$, since it is $U$, rather than $V$, that is one of the essential factors
to stabilize the spin-polarized edge states. We choose $U=1.2t$ in this
subsection. And the ZCNT's configuration is (6,0), which is metallic when the
Coulomb interaction is ignored.\cite{infinite}

\begin{figure}
[ptb]
\begin{center}
\includegraphics[
trim=0.000000in 0.568026in 0.000000in 0.000000in,
height=2.0505in,
width=3.1185in
]%
{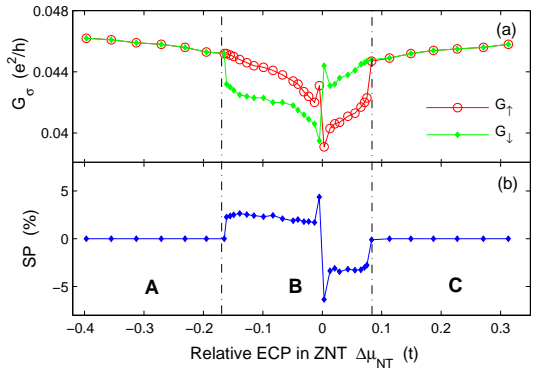}%
\caption{(Color online) (a) Spin-resolved conductance and (b) Spin
polarization of the conductance of the NM-ZCNT junction, as functions of the
relative electrochemical potential of the ZCNT. The division of the region A,
B and C is according to the magnitude of the SP.}%
\label{graph2}%
\end{center}
\end{figure}

Fig. \ref{graph2}(a) shows the spin-resolved conductance as a function of the
relative electrochemical potential of the ZCNT, defined by $\Delta\mu_{NT}%
=\mu_{NT}-$ $\mu_{NT}^{0}$, where $\mu_{NT}^{0}$ is the electrochemical
potential of the ZCNT in the half-filled case. And Fig. \ref{graph2}(b) shows
the spin polarization of the conductance, calculated from the data in Fig.
\ref{graph2}(a). We divide the electrochemical potential into three parts,
according to the spin polarization. In region A and C, the spin polarization
of the conductance is zero, while the spin polarization is not equal to zero
and even may change its sign in region B. In order to understand this
phenomenon, we investigate the energy levels of the edge of the ZCNT,
\textit{i.e.}, layer 1 in Fig. \ref{graph1}(a).

\begin{figure}
[ptb]
\begin{center}
\includegraphics[
height=2.3523in,
width=3.1194in
]%
{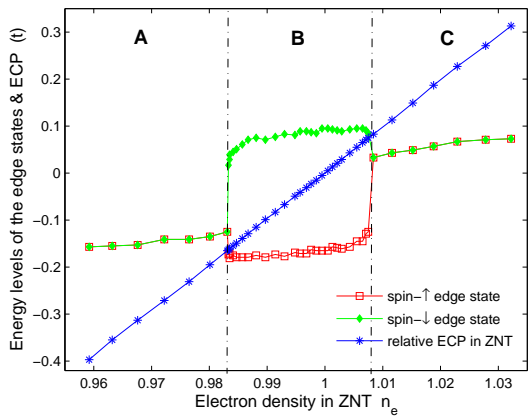}%
\caption{(Color online) The energy levels of the edge states and the relative
ECP in ZCNT as functions of electron density in the ZCNT. The whole range is
divided into three regions according to the relative magnitudes of the energy
levels of the edge states and the ECP of ZCNT, in accordance with Fig.
\ref{graph2}.}%
\label{graph3}%
\end{center}
\end{figure}

In Fig. \ref{graph3}, the square (diamond) markers denote the energy level of
the edge state with spin-$\uparrow$($\downarrow$), which are obtained by
locating the peaks of the spin-resolved local density of states (LDOS) in
layer 1. The abscissa $n_{e}$ is the electron density deep inside the ZCNT, by
our previously stated assumption it takes the value of the electron density at
the right side of the central ZCNT, or equally in the right ZCNT part. The
zero point of energy is chosen to be at $\mu_{NT}=\mu_{NT}^{0}$. The straight
line describes the dependence of the relative electrochemical potential
$\Delta\mu_{NT}$ on the electron density $n_{e}$. The linear relation between
them implies a constant density of states (DOS) and linear dispersion
relations around the Fermi level of the ZCNT. We again divide the whole range
into three parts according to the relative magnitudes of the energy levels of
the edge states and the electrochemical potential of the ZCNT. We notice that
this division is in accordance with Fig. \ref{graph2}, \textit{i.e.}, the
energy levels of the edge states overlap in region A and C, but split in
region B.%
%

\begin{figure}
[ptb]
\begin{center}
\includegraphics[
trim=0.000000in 0.568026in 0.000000in 0.000000in,
height=2.0513in,
width=3.1194in
]%
{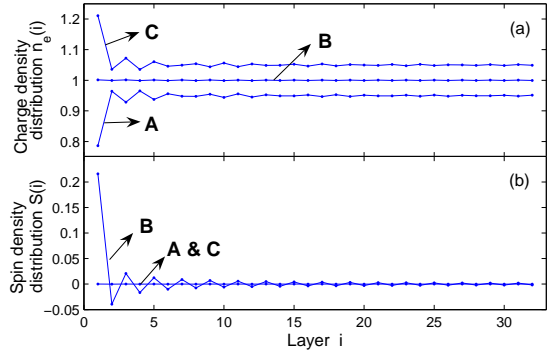}%
\caption{(Color online) (a) Electron's charge and (b) spin density
distribution along the ZCNT in the three configurations A, B and C. }%
\label{density}%
\end{center}
\end{figure}

In region A, $\Delta\mu_{NT}$ is below the energy levels of two edge states,
which are both empty. The charge density $n_{e}$ in layer 1 is very low as
shown in Fig. \ref{density}(a), and the spin density $S(i)$ is zero everywhere
as shown in Fig. \ref{density}(b). Thus the charge current is not spin
polarized through the junction in region A of Fig. \ref{graph2}. In region B,
$\Delta\mu_{NT}$ is between the energy levels of two edge states, thus only
one of them is occupied. This leads to a high spin density in layer 1 with its
charge density nearly equal to that in other layers as shown in Fig.
\ref{density}(a). The spin density oscillates in antiferromagnetic order in
the next several units and decays rapidly as shown in Fig. \ref{density}(b),
which is the same as expected in Refs.[\onlinecite{lieb1,lieb2}]. The
spin-polarized edge state acts as a spin-dependent barrier, and makes the
current through the junction spin-polarized in region B. In region C,
$\Delta\mu_{NT}$ is above the energy levels of both edge states, which are
occupied. The charge density in layer 1 is very high as shown in Fig.
\ref{density}(a), and the spin density is zero everywhere once again in Fig.
\ref{density}(b). So there's no spin-polarized current through the junction.

Furthermore, Fig. \ref{graph2}(b) shows an approximate antisymmetry, but not
an exact one. This fact reflects the partial particle-hole symmetry of the
Hubbard model on a biparticle lattice.\cite{lieb1,lieb2} As we can see in Fig.
\ref{graph1}, either the NM part or the ZCNT part is of biparticle lattice,
but the tunneling terms between them $H_{int}^{L}$ break the bipartite
structure of the whole system, and the symmetry is broken. However because of
a weak $H_{int}^{L}$, the SP still exhibits the antisymmetry approximately.

This subsection contains the main result of this paper, it describes three
phases of the edge of the ZCNT. They can emerge consequently by increasing the
electron densities or the electrochemical potentials of the ZCNT. As a result
the transition between these phases can be realized by adjusting the gate
voltage applied on the ZCNT in experiments.

\subsection{The case of $V\neq0$}

\begin{figure}
[ptb]
\begin{center}
\includegraphics[
height=2.3514in,
width=3.1185in
]%
{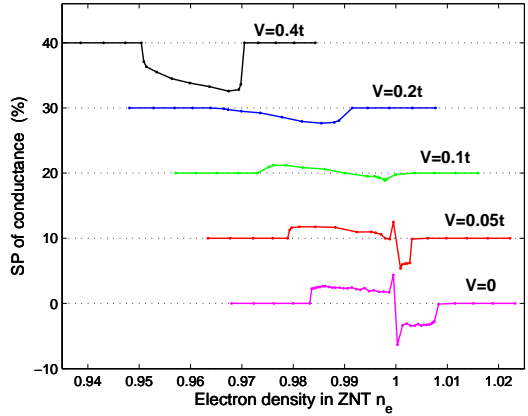}%
\caption{(Color online) Spin polarization of conductance as a function of
electron density in the ZCNT for different values of the nearest-neighbor
Coulomb interaction strength V, with vertical offsets of 10\% for clarity.}%
\label{graph5}%
\end{center}
\end{figure}

In this subsection, we investigate the effect of the nearest-neighbor Coulomb
interaction $V$. We take $U=2t$, and compare the spin polarization in the
cases of $V=0$, $0.05t$, $0.1t$, $0.2t$, and $0.4t$. Fig. \ref{graph5} shows
the SP of the conductance as a function of the electron density in the ZCNT,
which indicates that $V$ has two effects on the SP.

(i) When $V$ increases, the region of non-zero $S_{p}$ moves leftward. This
means that the energy levels of the edge states become lower from the middle
of the whole band as $V$ increases, which can be understood from the band
structure of the zigzag nanoribbons in Ref.[\onlinecite{stevennature}] because
the energy levels of the edge states of ZCNT can be obtained from the flat
bands of the zigzag nanoribbons by means of the band-folding
approach.\cite{infinite} The flat bands in Fig. 2(c) of
Ref.[\onlinecite{stevennature}] are lowered from the middle of the whole band,
and the same mechanism also happens in ZCNTs. Now we know it attributes to the
effect of the nearest-neighbor Coulomb interaction $V$.

(ii) When $V=0$, the spin polarization of the conductance is nearly
antisymmetric. The position of the sign change is about at $n_{e}=1$,
\textit{i.e.}, the half-filling point as shown in Fig. \ref{graph2}. This
property is relevant to the partial particle-hole symmetry of the Hamiltonian
Eq. (\ref{HNT}) at $V=0$,\cite{lieb1,lieb2} and the weak tunneling limit. But
the antisymmetry is broken as $V$ increases, the region of $S_{P}<0$ expands
leftward in Fig. \ref{graph5}. Thus we may conclude that in the cases with
$V=0.2t$, $0.4t$, as in Fig. \ref{graph5}, the lack of any sign change in the
spin polarization of the conductance is a result of the nearest-neighbor
Coulomb interaction $V$.

\subsection{Effect of the coupling strength}

\begin{figure}
[ptb]
\begin{center}
\includegraphics[
height=2.3514in,
width=3.1185in
]%
{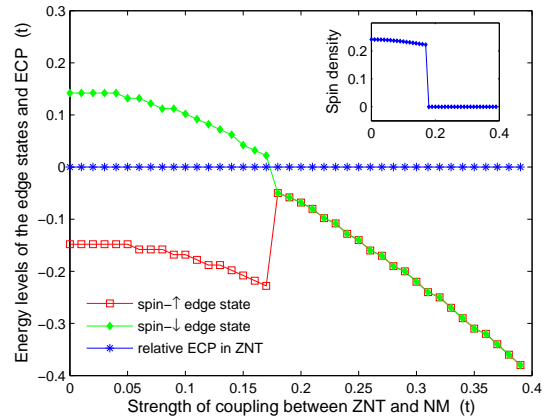}%
\caption{(Color online) The energy levels of the edge states and the relative
ECP in ZCNT as functions of the coupling strength between the ZCNT and NM
electrode. The inset shows the spin density on the edge of the ZCNT as a
function of the coupling strength. The parameters are $U=1.2t,$ $V=0.$}%
\label{4821}%
\end{center}
\end{figure}

In this subsection, we investigate the effect of the coupling between the ZCNT
and the NM part. Fig. \ref{4821} illustrates the energy levels of the edge
states and the relative ECP in the ZCNT as functions of the coupling strength
between the ZCNT and NM electrode. The relative ECP is always zero because the
ZCNT we study in this subsection is always half-filled. We can see the energy
levels of the edge states shift down when the strength of coupling increases.
There's a sudden change around $t_{int}^{L}=0.17t$, where the two energy
levels of the edge states are both occupied. This is a transition from the
single to double occupancy on the edge of the ZCNT. Because the edge states
are spin-polarized, the transition is also accompanied by a magnetic
transition at the edge. The inset shows the spin density on the edge of the
ZCNT as a function of the coupling strength $t_{int}^{L}$. When the edge
states are single-occupied, the edge of the ZCNT is magnetic. However, the
magnetism of the edge disappears suddenly at $t_{int}^{L}=0.17t$ when the two
edge states are both occupied, and the spin polarizations of the two states
cancel each other.

\section{Conclusion}

In summary, we have investigated the spin polarized transport in a junction of
a nonmagnetic metal lead and a zigzag carbon nanotube. Due to Coulomb
interaction of electrons, the zigzag carbon nanotube indicates strong
antiferromagnetic correlation at half-filling, which is enhanced near the
edge. The effect of the magnetic edge states leads to spin polarization of the
electronic transport.

Compared with the half-metallic zigzag graphene nanoribbons in Ref.
[\onlinecite{stevennature}], the NM-ZCNT junction device we present here can
be fabricated more easily because of the developed synthesis technique of
carbon nanotubes, though the spin polarization is weaker than that of the
zigzag nanoribbons, which can be 100\% because of their half-metallic
property. In both cases edge states are the keys in generating spin
polarization, but they act in different ways. The edge states of a nanoribbon
carry spin-polarized currents, while the edge state in our nonmagnetic
metal-zigzag carbon nanotube junction does not, actually it blocks the current
in a spin-dependent manner to produce spin polarization. The spin filter
function of a ZCNT hasn't been discovered in experiments so far. This is
possibly because the contact between the nanotube and the electrode has
allowed the current to enter the nanotube without crossing the edge, as shown
in Fig. 1 of Ref. [\onlinecite{experiment}]. This type of contact is called
\textquotedblleft side-contact\textquotedblright\ in Ref.
[\onlinecite{contact}]. What we need is the so-called \textquotedblleft
end-contact\textquotedblright, and the coupling between the ZCNT and the NM
electrode must be weak enough to preserve the edge state of the ZCNT. This way
the junction may act as a spin filter and produce spin-polarized tunneling
current.
\begin{figure}
[ptb]
\begin{center}
\includegraphics[
height=2.3514in,
width=3.1185in
]%
{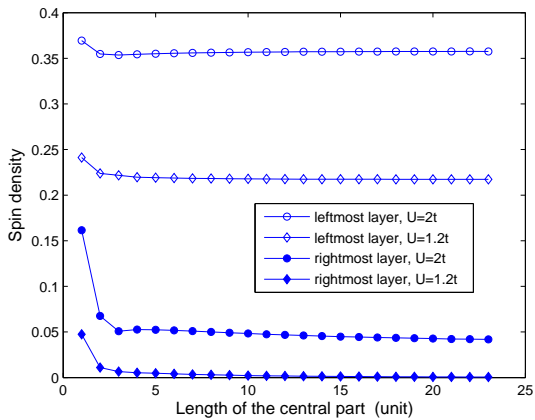}%
\caption{(Color online) Spin densities in the leftmost and rightmost layer of
the central part as functions of the lengh of the central part for $U=1.2t$
and $2t$, $V=0$. The spin density in the rightmost layer is multiplied by
(-1).}%
\label{spindensity}%
\end{center}
\end{figure}
\begin{figure}
[ptb]
\begin{center}
\includegraphics[
height=2.3514in,
width=3.1185in
]%
{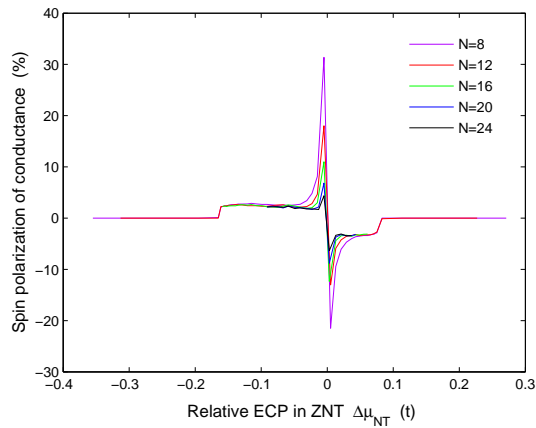}%
\caption{(Color online) Spin polarization of conductance as a function of the
relative electrochemical potential of ZCNT for different length of the central
part N=8,12,16,20,24.}%
\label{spn}%
\end{center}
\end{figure}

\begin{acknowledgments}
We thank Guan-Hua Chen and Fan Wang for useful discussions. This work was
supported by NSF of China under grant 90406017, 60525417, the NKBRSF of China
under Grant 2006CB921400, and the Research Grant Council of Hong Kong under
grant No.: HKU 7042/06 (S.Q.S).
\end{acknowledgments}

\section*{Appendix: Validity of the self-consistent method}

In the Sec. II(A), in order to use the technique of the principal
layer\cite{self1,self2} to calculate the Green's function of the semi-infinite
ZCNT lead, we divide the NM-ZCNT junction into three part: the left NM lead,
the central ZCNT part, and the right ZCNT lead. We take the left N units of
the ZCNT as the central part, and the rest of the ZCNT as the right
semi-infinite lead. It is assumed that the right ZCNT lead has the mean-field
Hamiltonian $h_{NT}^{R}$ with periodic distribution of local charge and spin
densities. This assumption becomes reasonable only when the length of the
central part N is large enough such that the edge effect vanishes completely
to the bulk properties. For a large N, the expectation of any physical
quantity in the ZCNT lead is assumed to be independent of N. We investigate
the spin densities in the leftmost and rightmost layers of the central part as
functions of N in Fig. \ref{spindensity}. It can be seen that the expectation
values converge quickly as N gets larger. The spin density in the leftmost
layer is large to form a magnetic edge state. The spin density in the
rightmost layer, which is multiplied by -1 in Fig. \ref{spindensity}, is
small, and converges to the amplitude of the spin density wave in an infinite
ZCNT\cite{HF1}. By the way, we have used the self-consistent Green's function
method on the infinite nanotubes, and recovered the results in Ref. [\onlinecite{HF1}].

Further more, we compare the SP of conductance for different length of the
central part, as shown in Fig. \ref{spn}. As N increases, most of the SP do
not change except the region of sign reversal. The peaks of SP become lower as
$N$ increases. Due to the restriction of our computer's ability, we just take
$N=24$ for the central part in this paper, and keep in mind the SP of
conductance we calculated may not be very accurate around the point of sign reversal.

We also make an error estimation of the self-consistent method. We run 10 more
iterations after the condition of convergence has been satisfied, and
calculate the relative standard deviation of $S_{p}$. The relative standard
deviation are all less than $1\%$ for dozens of random samples.

\end{document}